\documentclass[aps,prapplied,noshowpacs,noshowkeys,amsmath,amssymb,amsfonts,reprint,longbibliography]{revtex4-1}
\usepackage{graphicx}
\usepackage{dcolumn}
\usepackage{bm}
\usepackage{color}

\begin{document}
\title{Dipolar Rings of Microscopic Ellipsoids: Magnetic Manipulation and Cell Entrapment}
\author{Fernando Martinez-Pedrero$^{1,2}$}
\author{Andrejs Cebers$^{3}$}
\author{Pietro Tierno$^{1,2}$}
\email{ptierno@ub.edu}
\affiliation{
$^1$Departament de F\'isica de la Mat\`eria Condensada, Universitat de Barcelona, Barcelona, Spain.\\
$^2$Institut de Nanoci\'encia i Nanotecnologia, IN$^2$UB, Universitat de Barcelona, Barcelona, Spain.\\
$^3$Faculty of Physics and Mathematics, University of Latvia, Zellu 23, LV-1002, Riga.
}
\date{\today}
\begin{abstract}
We study the formation and dynamics of
dipolar rings composed by
microscopic ferromagnetic ellipsoids,
which self-assemble in water
by switching the direction of the applied field.
We show how to manipulate these fragile structures
and control their shape via application of external
static and oscillating magnetic fields.
We introduce a theoretical framework which describes
the ring deformation under an applied field,
allowing to understand the underlying
physical mechanism.
Our microscopic rings are finally used to capture,
entrap and later release a biological cell
via magnetic command,
i.e. performing a simple operation which can be
implemented in
other microfluidic
devices which make use of
ferromagnetic particles.
\end{abstract}
\maketitle
\section{Introduction}
A ring is a fascinating one-dimensional
topological structure
which encircles a finite two-dimensional
region of the space. When deformed,
a ring may reduce its enclosed surface area and
acquire a complex shape resulting
from competition between compression and
bending rigidity.
At the microscale, there are
different ways to assemble
colloidal particles into rings,
such as by using electric fields~\cite{Han03},
laser beams~\cite{Lut04,Nag14}, capillary forces~\cite{Yun11,Zha13}
or time-dependent magnetic fields~\cite{Erb09,Ray10,Sne11,Mar15}.
In contrast,
in the absence of external forces,
rings formed by spontaneous self-assembly
are more difficult to observe.
When the particles interact via isotropic and attractive
pair potentials,
the formation of compact clusters
is favored over open, ring-like structures~\cite{Str04}.
On the other hand,
anisotropic pair interactions
like those arising
from induced moments
promote chaining
along a defined direction~\cite{Skj83,Gou03}.
Dipolar rings have been observed with magnetized
spheres~\cite{Wen99,Van14} when balancing any external field,
and with Janus-magnetic microrods~\cite{Yan13},
when inverting the direction of the applied field.
Minimization of the system energy
under such conditions shows that
the equilibrium configuration of
dipolar particles corresponds to
closed rings~\cite{Gen70,Wen99,Pro09}.
While the formation of dipolar
rings~\cite{Tav99,Kun01,Mor03,Gha03,Dun04,Sne05,Pro11,Tav12,Rov12,Mes14,Mes15}
and their rupture under an external field~\cite{Kun01,Yan13}
have received attention
in the past, no work demonstrates
the possibility to manipulate
these fragile assembly
without breaking them,
and in such a way to perform
functional tasks.

Here we
study the formation and
dynamics of dipolar rings self-assembled from
microscopic hematite ellipsoids,
ferromagnetic and anisotropic particles with a
magnetic moment perpendicular to their long axis.
We manipulate
these self-assembled structures by using an external field
and present a theoretical framework
that explains the ring
deformation and dynamics
under the applied field.
Finally, we use these rings to entrap and later release
micro-objects such as a biological cell,
in a
fluidic environment,
extending the potentiality of our technique
to applications in fluidic microsystems
based on precise micro-encapsulation.

\section{Experimental method and ring formation}

We synthesize monodisperse hematite ellipsoids following
the method developed by Sugimoto {\it et al.}~\cite{Sug93}. Inspection of
scanning electron microscopy images of the particles reveals that they
are prolate ellipsoids with major (minor) axis
equal to $a=1.80 \, {\rm \mu m}$ ($b=1.33 \, {\rm \mu m}$ resp.).
After synthesis, the ellipsoids are
dispersed in pure water (milliQ, Millipore),
stabilized with sodium dodecyl sulfate and the pH
is adjusted to $8.5-9.5$.
The particles sediment close to a glass plate,
where they remain quasi two-dimensional confined due to gravity.
To visualize the particle dynamics, we use a upright light microscope (Eclipse Ni,
Nikon), and record their position with a CCD
camera (Scout scA640-74f, Basler) working at 50 fps.
We apply different external magnetic fields by using a custom-made
triaxial coil system connected either to a power supply (EL 302RT, TTi),
when needed to generate a DC field, or
to a wave generator (TTi-TGA1244, TTi) feeding a power amplifier
(IMG STA- 800, Stage Line) when needed to generate AC magnetic fields~\cite{Tie09}.

\begin{figure}[t]
\begin{center}
\includegraphics[width=\columnwidth,keepaspectratio]{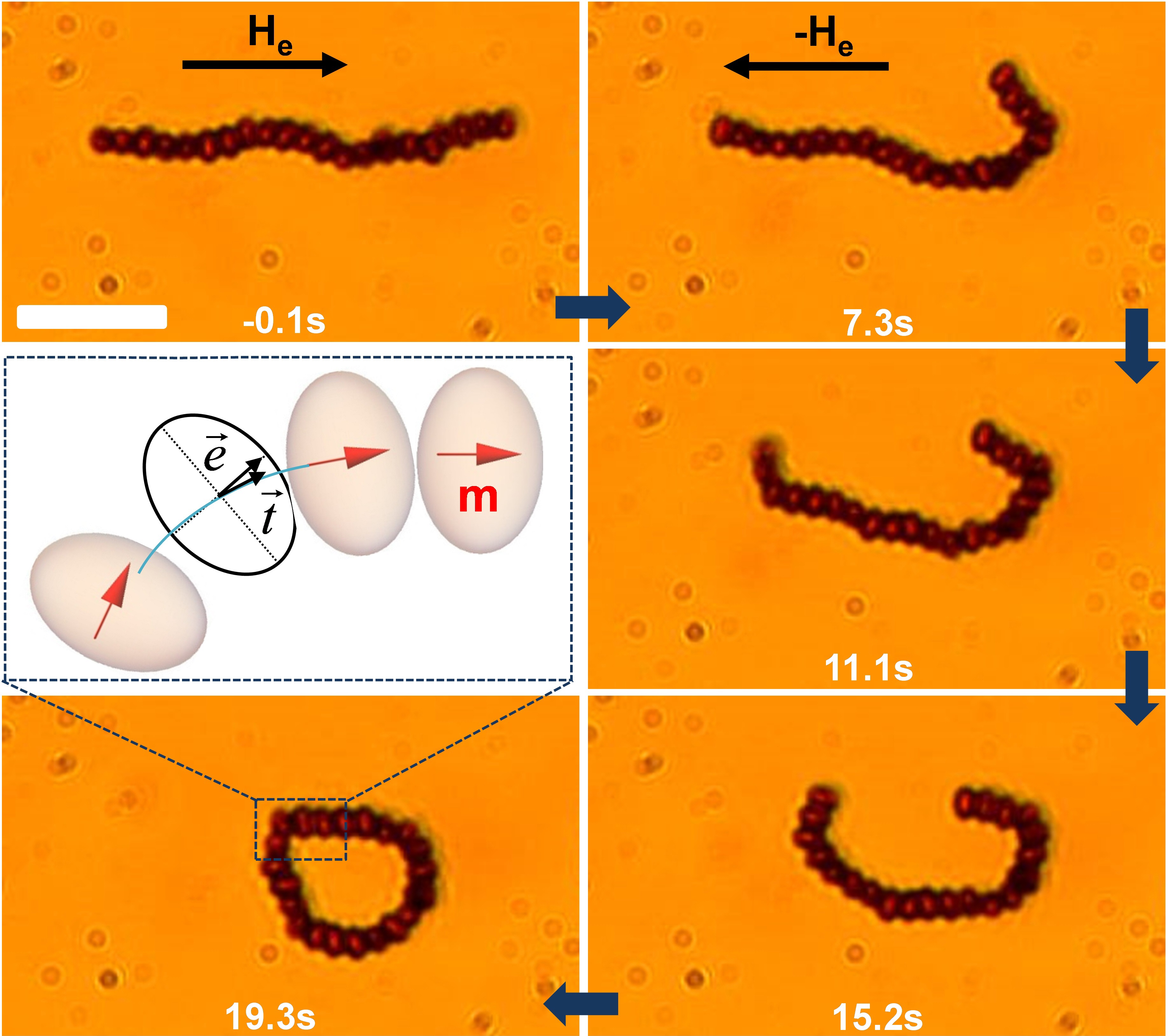}
\caption{Sequence of images showing
the formation of a colloidal ring composed
of $N=21$ hematite ellipsoids.
The initial image shows a ribbon oriented along the
direction of the earth
magnetic field $H_e$. At $t=0$,
a static field $H=-2H_e$
is applied and
the ribbon closes into a ring after $19.3 {\rm s}$ (see Video 1).
Scale bar is $10 {\rm \mu m}$.
The schematic in the middle left
illustrates four ellipsoids
with magnetic moments $\vec{m}$
perpendicular to their long axis
and the vectors $\vec{t}$ and $\vec{e}$.}
\label{fig_1}
\end{center}
\end{figure}

In the absence of
any applied field, the ellipsoids rapidly assemble into
a ribbon-like structure due
to the presence of a small
permanent moment $\vec{m}$ in the
particles,
perpendicular to the ellipsoid long axis.
Once formed, the ribbon does not orient
randomly, but aligns in average along
the direction given by the earth magnetic field,
$H_e \sim 0.5 {\rm Oe}$.
As shown in Fig.1, by applying a small field
of amplitude $H=2H_e$ in
the opposite direction,
we are able to form a dipolar
ring from a ribbon configuration, see Video 1.
\begin{figure}[t]
\begin{center}
\includegraphics[width=0.9\columnwidth,keepaspectratio]{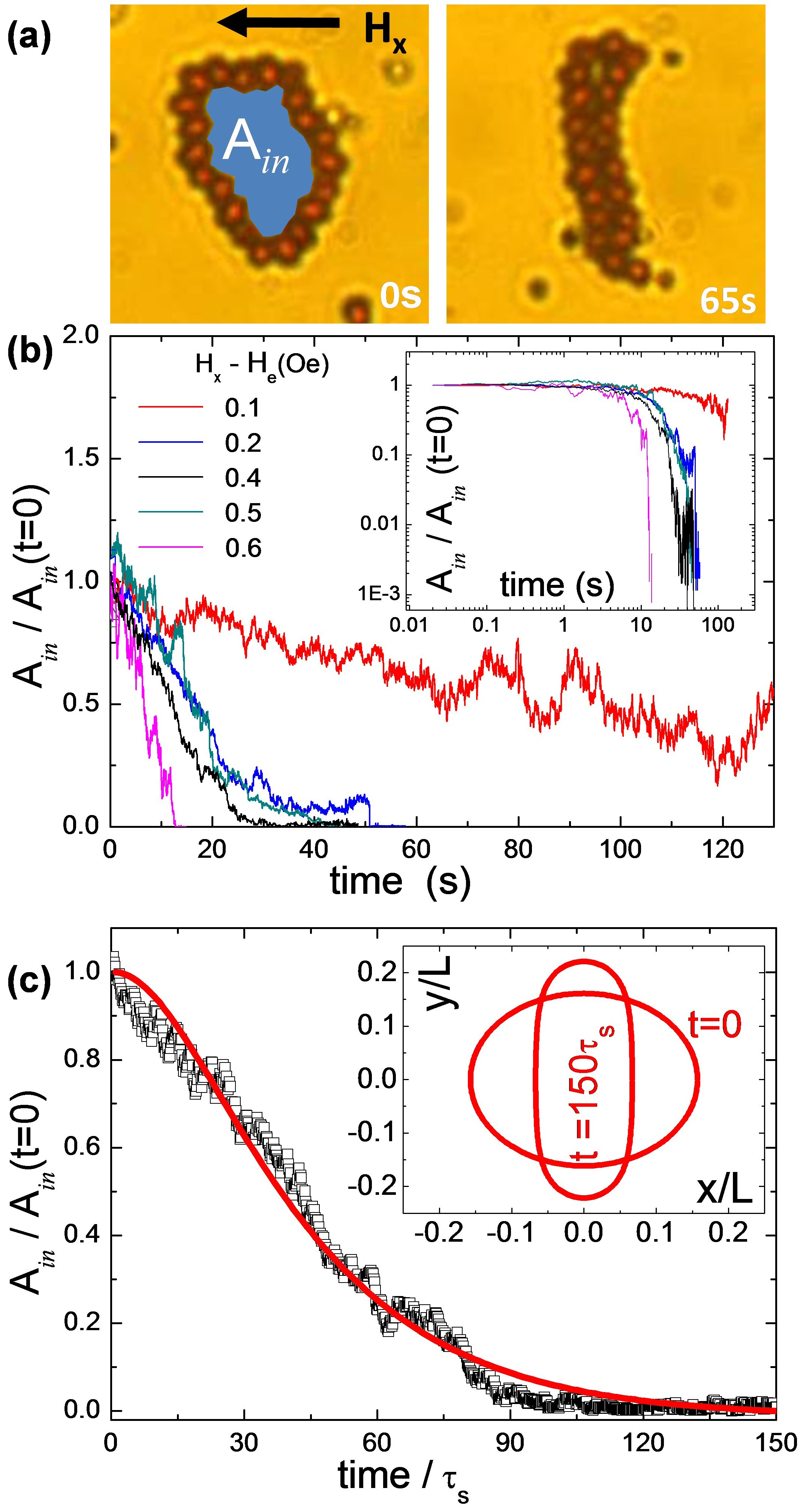}
\caption{(a)
Deformation of a colloidal ring composed by $N=20$
ellipsoids when subjected to a static field $H_x$.
See also Video 2 that shows
a smaller ring
in the proximity of a ribbon.
(b) Normalized area enclosed by the ring versus
time for different applied magnetic fields.
$A_{in}$ denotes the internal area of the ring.
Inset shows the same plot but in a log-log scale.
(c) Normalized area versus re-scaled time $t/\tau_s$
($\tau_s=3.5 {\rm s}$)
for a dipolar ring of $N=20$ particles
compressed via application of a
constant field $H_x - H_e = 0.4 {\rm Oe}$.
Scattered points are experimental data,
while continuous red line is the result of the
numeric integration of
Eq.6.
Inset shows the calculated ring contour at different
rescaled times.}
\label{fig_2}
\end{center}
\end{figure}
The ribbon slowly deforms assuming an hairpin-like
shape, and its two free ends approach each other,
and finally join
forming a ring.
The kinetic pathway described in Fig.1 is only
one possible route towards ring formation.
We also find
that after applying the small external field $H$,
the ribbon can break into two
rings via an S-shape
deformation, when the two
ends of the chain curve in opposite directions,
as observed with ferromagnetic Janus rods~\cite{Yan13}.

\section{Analysis of ring deformation}

We control the shape of the ring
by further imposing an additional static field
in the particle plane.
As shown in Fig.2(a),
the ring shrinks
along
the applied field. In contrast, the
ribbons orient along the applied field,
as shown in Video 2.
The applied field forces the
magnetic moments of the particles
to deviate from
the tangent direction of the
ring, described by the vector $\vec{t}$. Thus, the magnetic energy will depend
upon the angle between $\vec{t}$ and the applied field, $\vec{H}=H\vec{h}$.
Given the small anisotropy of our hematite particles, with an aspect ratio of $\sim 0.74$, 
we describe the ring as an ensemble of interacting 
spheres characterized by a point dipole at the center.
We write the total magnetic energy per unit length of the ring as
\begin{equation}
e_{m}=-M\vec{e}\cdot\vec{H}-\frac{1}{2}\Delta N M^{2}(\vec{e}\cdot\vec{t})^{2}  \, \, ,
\label{Eq:1}
\end{equation}
being $M$ the magnetization per unit length,
$\vec{e}$ the unit vector along the
local particle magnetization
which may vary along the contour of the ring,
$\Delta N=6 K_{N}'/d^2$ the demagnetization factor,
$d$ the particle diameter
and $K_{N}'=\frac{1}{4}\sin^{3}{(\pi/N)}\sum^{N}_{i=2}\frac{1-\cos{(2\pi(i-1)/N)}}{\sin^{5}{(\pi(i-1)/N)}}$,
see the Appendix A for a detailed derivation.
We note here that the 
particle diameter $d$ used in the model is
replaced in our calculations by the minor axis of the ellipsoid $b$.
As demonstrated in the Appendix, Eq.1 is rather general, and can be equally applied to
both a chain or a ring of dipoles.
In the limit of small applied fields,
Eq.1 can be transformed (see Appendix B) to:
\begin{equation}
e_{m}=-MH\vec{t}\cdot\vec{h}+\frac{H^{2}}{2\Delta N}(\vec{t}\cdot\vec{h})^{2} \, \, .
\label{Eq:2}
\end{equation}
Since $\vec{t}=d\vec{r}/dl$,
the first term in Eq.2 vanishes after integration along the
closed centerline of the ring and only the second term 
gives a non-vanishing contribution to the energy of the ring.
Further, in the limit of small fields,
the ring has a negative anisotropic magnetic susceptibility
$\chi_{\parallel}-\chi_{\perp}=-1/\Delta N$  
($\chi_{\parallel},\chi_{\perp}$ denote parallel and perpendicular to the local tangent direction respectively).
Therefore there is a tendency to increase the
length of the elements of the ring which are perpendicular to the external  field.
Considering the ring in the frame of the Kirchoff model of elastic rods, with added
magnetic energy terms~\cite{Cebers1},
the torque balance equation per unit length of the ring reads
\begin{equation}
\frac{d\vec{T}}{dl}+[\vec{t}\times\vec{F}]+MH[\vec{t}\times\vec{h}]-\frac{H^{2}}{\Delta N}(\vec{t}\cdot\vec{h})[\vec{t}\times\vec{h}]=0 \, ,
\label{Eq:3}
\end{equation}
where $\vec{T}$ and $\vec{F}$ are the torque and the force
in the cross-section of the ring respectively.
The total force is given by,
$\vec{F}=\vec{F}_{e}+\vec{F}_{m}$.
Here $\vec{F}_{e}$ includes the
contribution of the elasticity,
\begin{equation}
\vec{F}_{e}=-A\frac{d^3\vec{r}}{dl^3}+\Lambda\frac{d\vec{r}}{dl} \, \, ,
\label{Eq:Fe}
\end{equation}
where $A$ is the bending modulus,
the Lagrange multiplier $\Lambda$
accounts for inextensibility of the rod~\cite{Cebers2},
and the magnetic shearing force is given by
\begin{equation}
\vec{F}_{m}=-MH\vec{h}+\frac{H^{2}}{\Delta N}(\vec{t}\cdot\vec{h})\vec{h} \, \, .
\label{Eq:4}
\end{equation}
Since in Eq.5
the first term is homogeneous, it does not enter in the equations of motion.
The equation of motion for the ring
reads:
\begin{equation}
\zeta_e\vec{v}=\frac{d\vec{F}_{e}}{dl}+\frac{H^{2}}{\Delta N}\vec{h}\Bigl(\frac{d^{2}\vec{r}}{dl^{2}}\cdot\vec{h}\Bigr) \, \, ,
\label{Eq:5}
\end{equation}
where
$\vec{v}$ denotes the velocity of the element of the ring,
$\zeta_e=4\pi\eta$ is the isotropic hydrodynamic drag coefficient
of the ring per unit contour length
and $\eta$ the viscosity of water.
The total energy functional is given by:
\begin{equation}
E=\frac{A}{2}\int{\Bigl( \frac{d\vartheta}{d l}\Bigl)^2 dl} +\frac{H^2}{2\Delta N}\int{\cos^2{\vartheta} dl}  \, \, ,
\label{Eq:6}
\end{equation}
being $l$ the natural parameter along the ring contour,
$\vec{t}=(\cos{\vartheta},\sin{\vartheta})$
and $d\vartheta/dl=-1/R$ the ring curvature.
Minimizing Eq.7 gives the Euler-Lagrange equation
\begin{equation}
A\frac{d^2\vartheta}{dl^2}+\frac{H^2}{\Delta N}\cos{\vartheta}\sin{\vartheta}=0 \, \, ,
\label{Eq:7}
\end{equation}
written assuming an inextensible ring.
Starting from this equation, it can be shown (Appendix C) that the shape
of the ring is described by the
set of equations (here, the contour length $l$ is scaled with the total length of the ring $L$):
\begin{equation}
\frac{dx}{dl}=cn\Bigl( \sqrt{\frac{C_m}{k^2}l},k \Bigl); \, \, \, \,
\frac{dy}{dl}=sn\Bigl( \sqrt{\frac{C_m}{k^2}l},k \Bigl) \, \, ,
\label{Eq:8}
\end{equation}
written in a dimensionless form. Here
$cn(u;k)$ and $sn(u;k)$
are the Jacobi elliptic functions of the first and second kind,
$u$ is the argument and $k$
the modulus,
$k^2=H^2/(2 \Delta N C_1)$, $C_1$
is the first integral of Eq.8
and $C_m$ is the magnetoelastic number defined as
\begin{equation}
C_m=\Bigl(\frac{L}{d}\Bigr)^{2}\frac{H^{2}}{(m/d^{3})^{2}}\frac{1}{3K_{N}'} \, \, ,
\label{Eq:9}
\end{equation}
where the bending modulus of the ring is only given
by the magnetic dipolar interactions
according to $A=m^{2}/(2 b^{2})$ \cite{Ceb05}.
The shape of the ring can be calculated
by numerical integration of Eq.6.

\section{Experimental results}

In order to validate our theoretical model, 
we have performed a series of experiments 
by measuring the area enclosed 
by a ring 
$A_{in}$ as it reduces due to the
applied field $H_x$,
at different field amplitudes. 
These realizations 
are shown in Fig.2(b), which reports 
the normalized internal area 
of the rings $A_{in}/A_{in}(t=0)$ 
as it reduces with time. 
\begin{figure}[t]
\begin{center}
\includegraphics[width=\columnwidth]{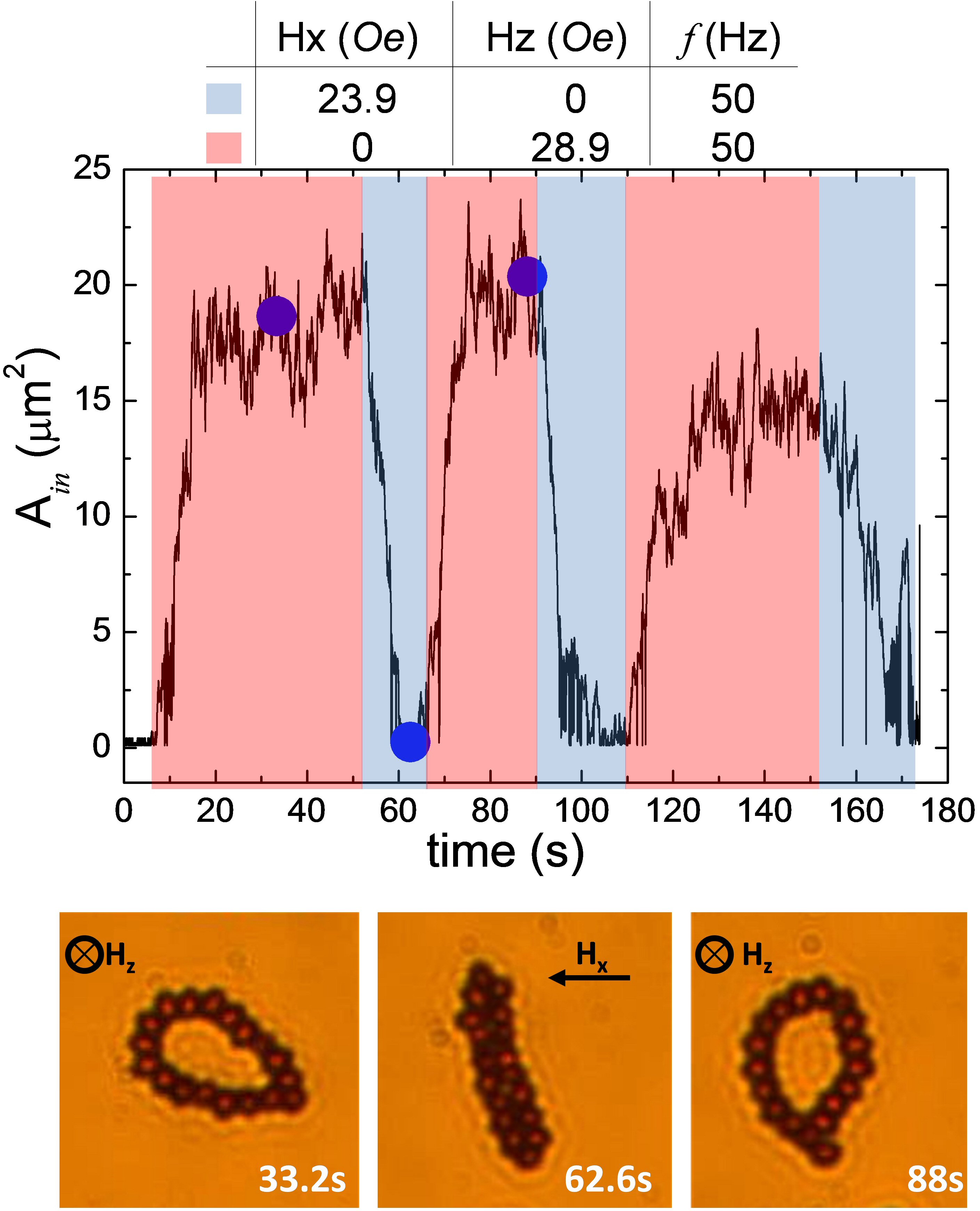}
\caption{Time evolution of the area $A_{in}$
of a dipolar ring composed of $N=18$ particles
and subjected to
a $50 \, {\rm Hz}$ oscillating
magnetic field applied either
perpendicular to the ring ($H_z$)
or in the plane of the ring ($H_x$).
The filled scattered points indicate
the location of the snapshots displayed
at the bottom,
which show the
ring deformations.}
\label{fig_3}
\end{center}
\end{figure}
A minimum field of $H_x-H_e=0.1 {\rm Oe}$
is required to compress the ring.
For field strengths higher than
$H_x-H_e=0.2 {\rm Oe}$, we find that ring area
reduces within $60s$.
In the range of values
between $0.2$ and $0.5 {\rm Oe}$,
the dynamics of the ring deformation 
is difficult to distinguish, 
as shown in the inset in Fig.2(b). 
Thermal fluctuations make difficult 
to resolve  the small differences 
in the shrinking dynamics of the
rings, in the narrow range of applied fields.
At fields larger than $H_x-He=0.6 {\rm Oe}$,
the ring structures break during the
shrinking process.
In Fig.2(c) we compare the
experimental data for one field $H_x-H_e=0.4 {\rm Oe}$
with the theoretical predictions,
and rescaled the time by
$\tau_s=\tau_e \cdot 10^{-6} {\rm s}$,
where $\tau_e$
is the characteristic elastic
deformation time
used as adjustable parameter.
The continuous red line fitting 
the experimental data in Fig.2(c) 
is obtained by numerically solving 
Eq.6, following the numerical approach 
described in~\cite{Cebers2}. The equilibrium 
shapes calculated according to Eqs.9 coincide
with the shape obtained from the numerical integration of Eq.6.
We use a value of $C_m=557.7$
assuming $K_{N}' \sim 1$, $d=b$
and a permanent moment
of the ellipsoid
given by $m=2.3\cdot 10^{-13} \, {\rm emu}$,
as previously measured from the reorientation dynamics
of an individual ellipsoid~\cite{Fer16}.
The good agreement with the experimental data
reported in Fig.2(c) is obtained
using a value of $\tau_e=3.5 \cdot 10^6 {\rm s}$.
On the other hand,
we can independently estimate this parameter,
by assuming that the bending modulus is
given only by dipolar interactions,
$A=m^2/2b^2=1.5 \cdot 10^{-18} \, {\rm erg \cdot cm}$.
We thus find a characteristic timescale
given by $\tau_e=\zeta_e L^4/A= 4.2\cdot 10^6 \, {\rm s}$,
which is very close to the experimental value.
Moreover, the persistence length
of the chain can be estimated as the ratio
$l_p=A/k_BT\cong 0.3 \, \mu m$,
being $k_B$ the Boltzmann constant and $T=293{\rm K}$
the experimental temperature.
The small value of the persistence length obtained, smaller than the particle's size, may result 
from neglecting other attractive forces that increase the ring compactness. 
A small persistence length 
is consistent with the experimental observation that 
these rings are rather fragile, 
and easily deform
due to thermal fluctuations.
Effectively,
we find that the
these rings can easily break
when the applied constant field exceeds the value
of $\sim 0.6~{\rm  Oe}$.

The self-assembled microscopic rings
studied here are kept only by dipolar forces,
thus they can be easily broken by small static fields.
However, we find that
an oscillating field
applied perpendicular to the plane of the
ring, is able to stabilize
the ring shape even having a much large amplitude.
An oscillating field can be used to
open a ring that has been previously closed,
thus making completely reversible the induced
\begin{figure*}[th]
\begin{center}
\includegraphics[width=0.9\textwidth,keepaspectratio]{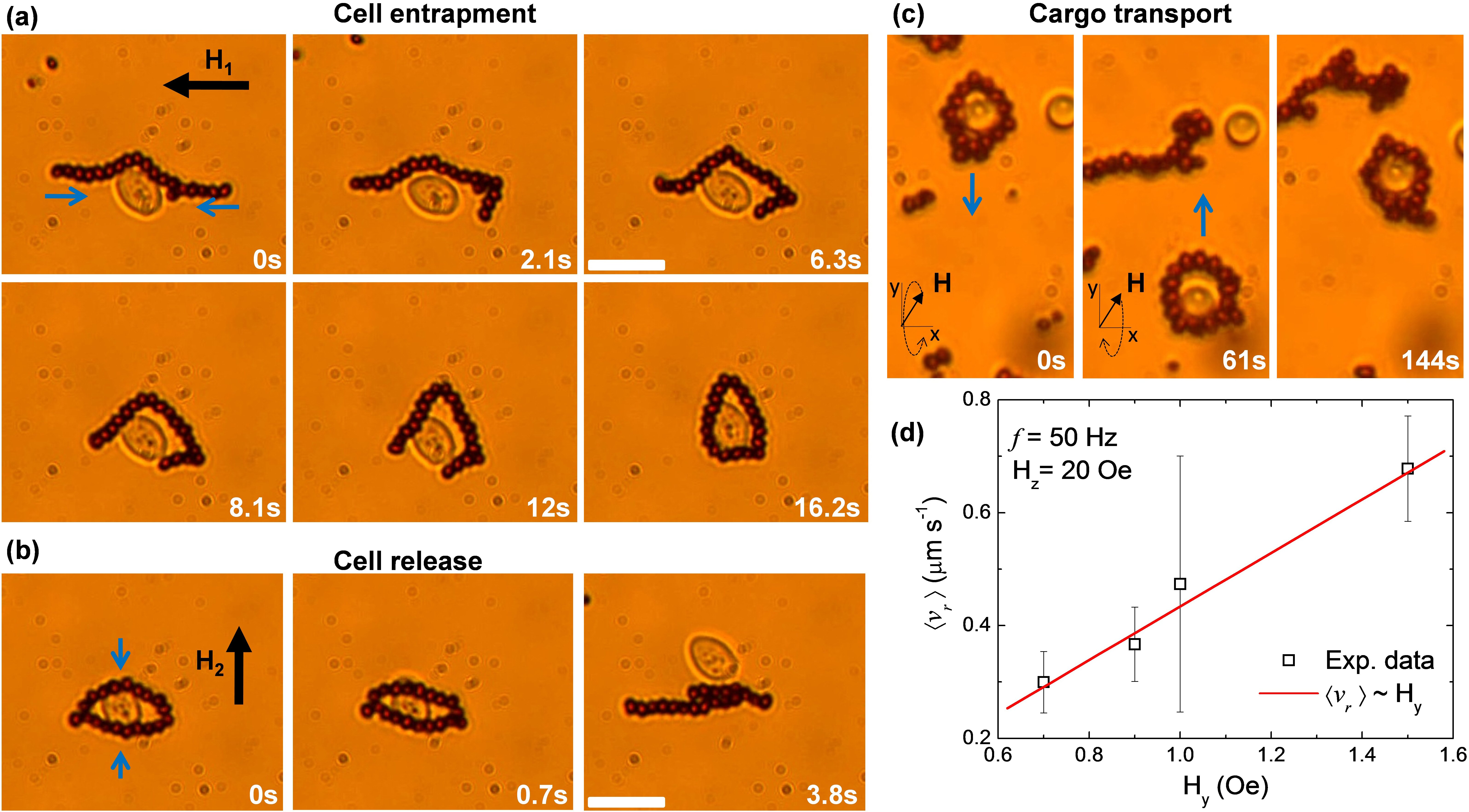}
\caption{(a) Sequence of images showing the
entrapment of a yeast cell
by a ribbon composed of $19$ ellipsoids.
Under a static field with amplitude $H_1= -2H_e$
the ribbon encircles the cell after $16.2s$ (Video 4).
(b) Images showing the release of the cell
upon application of a static field $H_2= 0.2 {\rm Oe}$
which compresses the colloidal ring (Video 5).
Scale bars are $10 \, {\rm \mu m}$ in both sequences.
(c) Transport of a $3 {\rm \mu m}$ SiO$_2$
particle entrapped in a ring via 
application of a precessing magnetic field 
$\vec{H}\equiv (H_x, H_y \cos{(2\pi f t)}, H_z \sin{(2\pi f t)})$
with frequency $f=20 {\rm Hz}$
and amplitudes $H_x-H_e=0.1 {\rm Oe}$, $H_y=0.7 {\rm Oe}$ and $H_z=7.1 {\rm Oe}$.
After $t=61 {\rm s}$ the field polarity
is inverted, and the ring changes its direction of motion
(Video 6).
(d) Average speed of the ring $\langle v_r \rangle$
versus amplitude of the in-plane field $H_y$ for a ring transporting
the silica cargo. The continuous line denotes 
a liner fit to the experimental data.}
\label{fig_4}
\end{center}
\end{figure*}
deformation. These features
are shown in Fig.3, which displays the controlled
compression and
expansion of one colloidal ring subjected to
an oscillating magnetic field
applied perpendicular to the
ring, $H_z$ (expansion),
or in the ring plane, $H_x$ (compression).
We note here that the expansion of a dipolar
ring cannot be
achieved by using a simple static
field
since the latter will
force the magnetic moments of the ellipsoids
to orient along the field.
In this situation,
all moments in the rings will be
perpendicular to the center-to-center distance, 
giving rise to a net repulsive dipolar interaction between them.
In contrast,
for an applied field oscillating faster than
the individual
reorientation of the ellipsoids, the
particle moment is unable to follow the direction of the field.
As a consequence, the moments orient
perpendicular to the direction of
the applied field~\cite{Bel06,Erg09,Fer16}.
Thus, for an oscillating field perpendicular 
to the ring plane,
the moments
will point along the plane of the ring,
a situation which is favourable to the ring formation,
similar to the configuration
shown in Fig.1. 

\section{Cargo entrapment and transport.}

Beside the interest in studying self-assembly
processes of magnetic particles~\cite{Yan13},
dipolar rings can also be used for
lab-on-a-chip operations, such as
to entrap  and later release
microscopic objects
via external command.
We demonstrate both features in Fig.4
(Video 3- Video 5),
where a ribbon 
forms a ring which
entraps one yeast cell within its boundaries.
Before the trapping, and in order to approach the cell, the ribbon was
translated at a constant speed of $\langle v \rangle = 0.6 \mu m s^{-1}$
upon application of an external precessing magnetic field~\cite{Tie07}
given by, $\vec{H}\equiv(H_x, H_y \cos{(2\pi f t)}, H_z \sin{(2 \pi f t)})$,
with frequency $f= 50 {\rm Hz}$,
and amplitudes $H_x-H_e=0.3 {\rm Oe}$, $H_y=27.4 {\rm Oe}$ and $H_z=3.5 {\rm Oe}$.
The applied field performs a conical precession around
the $x-$axis;
the static component $H_x$
and the oscillating component $H_y$
orient the ribbon along the $x-$direction and
keep the
particles attached to the ribbon.
The rotating component induces the rotational motion
of the individual ellipsoids.
The ribbon thus translates
in the perpendicular ($y-$)direction,
due to the hydrodynamic coupling
of the rotating particles
with the close glass surface~\cite{Tie08}.
This mechanism is similar to the propulsion of
ensemble of paramagnetic colloids
recently reported by Casic {\it et al.}~\cite{Cas13}.
Using this strategy,
the ribbon is transported close to the cell
from any location of the plane, see Video 3.
After that, the precessing field
is switched off, and a static field
$H_x= 0.7{\rm Oe}$, applied contrary
to $H_e$, is used to form a ring which complete encircles
the cell after $16.2 {\rm s}$, Fig.4(a) (Video 4).
Once captured, the cell
performs small thermal fluctuations within
the magnetic ring, but it
remains stably confined over time.
In order to release the cell,
without dismantling the magnetic structure,
we compress the ring by applying
a static magnetic field
$H_2= 0.2{\rm Oe}$ along the $y-$direction,
Fig.4(b) (Video 5)
The ring reduces its size and after $0.7s$
the cell is squeezed off from the top part
to be finally completely expelled from
the ring.
This operation is reversible,
since the ribbon of particles can be
transported again close to another microscopic object,
where it can be transformed back
into a ring
to encircle it.
When all the microscopic entities are 
two-dimensionally confined by gravity 
on the glass surface, the presence of 
the ring  prevents future encounters between 
the cell and its neighbors.  However, the cell 
is not protected in the third dimension. Our 
dipolar rings can be used to create a "protective" 
environment  around a biological object when the 
system is two-dimensionally confined.

While the capture and the release
have been all
made by using uniform external fields,
we have found that the rings can be also
slowly transported by applying
rotating fields,
a further functionality which could be employed to translate
micro-objects entrapped within this structure.
In particular, in Fig.4(c) we show
the transport of a $3 {\rm \mu m}$ SiO$_2$ 
particle which was previously entrapped in a 
ring composed of $N=15$ ellipsoids. 
The ring propulsion was induced by applying 
an external rotating magnetic field in the 
$(y,z)$ plane, perpendicular to the static field $H_x$. 
As shown in the corresponding Video 6,
the rotating field forces the ellipsoids to perform a
rotational motion around their short axis. 
Similar to the translation of the linear ribbon, 
the mechanism of motion of the ring results from 
the rotational movement of the composing ellipsoids 
close to the glass substrate. During the motion of the ring,
the trapped particle is unable to escape from the ring 
boundaries, and it is transported along the glass plate 
at a constant average speed $\langle v_r \rangle= 0.59 {\rm \mu m^{-1} s}$. 
The transport process occurs without the dis-assembly 
of the ring structure during its motion,
and reverting the polarity of the rotating
field allows us to change direction of motion. Moreover,
we find that the ring speed can be 
well controlled by tuning the amplitude of the 
in-plane field $H_y$, Fig.4(d).
In the range of the field strengths 
here employed, we find that $\langle v_r \rangle$ scales linearly 
with $H_y$, reaching a speed up to $0.7  \mu m s^{-1}$ for $H_y=1.5 {\rm Oe}$. 
Compared to the propulsion of the linear ribbon, 
and at parity of applied field conditions, 
the average speed is smaller since the ring 
curvature minimizes the rotational motion 
of the particles located close to the ring equator.

\section{Conclusion}

In summary, we have studied
the assembly and
controlled deformation of dipolar rings under external magnetic fields.
While in a previous article~\cite{Fer16}
the  ring formation was reported, we
demonstrate here real time manipulation of
these fragile structures, and develop a
theoretical model which explains the ring
compression along the direction of the applied field.
Our microrings are composed by
synthesized hematite ellipsoids,
similar self-assembled structures have been observed with other
ferromagnetic particles on different length scales~\cite{Kun01,Gha03,His09,Wei11}.
Finally, the possibility to use these ellipsoids 
to entrap microscopic objects, like biological cells,
shows the potentiality of this method
for controlled encapsulation
in fluidic environments.

\acknowledgments{Acknowledgments}
F. M.P. and P. T. acknowledge support from the European Research
Council (Project No. 335040). P. T. acknowledges support from Mineco
(No. RYC-2011-07605 and No. FIS2013-41144-P) and
AGAUR (Grant No. 2014SGR878).
A. C. acknowledges support
from National Research Programme No. 2014.10-4/VPP-
3/21.

\clearpage

\appendix
\section{Demonstration that Eq.1 holds for dipolar chain and rings}
Let us first consider the case of a chain of $N$ dipoles.
Its dipolar energy reads
\begin{equation}
U_{chain}=\sum^{N-1}_{i=1}\sum^{N}_{j=i+1}\Bigl(\frac{\vec{m}_{i}\vec{m}_{j}}{r^{3}_{ij}}-\frac{3(\vec{r}_{ij}\cdot\vec{m}_{i})(\vec{r}_{ij}\cdot\vec{m}_{j})}{r_{ij}^{5}}\Bigr) \, \, ,
\label{Eq:1}
\end{equation}
where $\vec{r}_{ij}=d(j-i)\vec{t};\vec{m}=m(e_{x},0,e_{z});~\vec{t}=(0,0,1)$, in the case when magnetic moment deviates from the axis of the chain. Then
\begin{equation}
U_{chain}=\frac{m^{2}}{d^{3}}\sum^{N-1}_{i=1}\sum^{N}_{j=i+1}\frac{1}{(j-i)^{3}}(1-3(\vec{e}\cdot\vec{t})^{2}) \, \, .
\label{Eq:2}
\end{equation}
When $\vec{e}\cdot\vec{t}=1$, Eq.(\ref{Eq:2}) transforms in 
the well-known relation for the energy
of a straight chain
\begin{equation}
U^{0}_{chain}=-\frac{2m^{2}}{d^{3}}NK_{N} \, \, ,
\label{Eq:3}
\end{equation}
where
\begin{equation}
K_{N}=\frac{1}{N}\sum^{N}_{l=1}\frac{N-l}{l^{3}}  \, \, .
\label{Eq:4}
\end{equation}
To consider the deformation of the chain in an applied external
field, the only important part 
in Eq.A2 is the one which depends on the orientation of the magnetic moment with
respect to the axis of the chain:
\begin{equation}
U_{chain}=const-\frac{3m^{2}}{d^{3}}NK_{N}(\vec{e}\cdot\vec{t})^{2} \, \, .
\label{Eq:5}
\end{equation}
Since $M=m/d$ is the magnetization per unit length and $Nd$ is the length of the chain, 
we have for the dipolar energy per unit length:
\begin{equation}
U_{chain}=-\frac{3M^{2}K_{N}}{d^{2}}(\vec{e}\cdot\vec{t})^{2} \, \, ,
\label{Eq:6}
\end{equation}
which is the second term of the right hand side of the Eq.1 of the paper.
Note that in Eq.(A6) the constant term is
omitted since the configuration of the 
ring is only determined by the orientation dependent term.

With more effort, we arrive at to a similar expression for the case of the ring.
The positions of the particles in a ring with a radius $R$ are
\begin{equation}
\vec{r}_{i}=R(\cos{(2\pi(i-1)/N)},\sin{(2\pi(i-1)/N)})  \, \, ,
\label{Eq:7}
\end{equation}
The magnetic moments form a constant angle
with the azimuthal direction $\vec{e}_{\varphi}(i)=(-\sin{(2\pi(i-1)/N)},\cos{(2\pi(i-1)/N)})$, 
where $\vec{e}_{i}=t_{r}\vec{e}_{r}(i)+t_{\varphi}\vec{e}_{\varphi}(i)$.
For the dipolar energy per particle we have
\begin{equation}
U_{1}=\sum^{N}_{i=2}m^{2}\Bigl(\frac{\vec{e}_{i}\cdot\vec{e}_{1}}{r^{3}_{i1}}-\frac{3(\vec{r}_{i1}\cdot\vec{e_{i}})(\vec{r}_{i1}\cdot\vec{e}_{1})}{r^{5}_{i1}}\Bigr) \, \, .
\label{Eq:8}
\end{equation}
Since the magnetic  moments are turned by the  same angle $\vec{e}_{i}\cdot\vec{e}_{1}=\vec{e}_{\varphi}(i)\cdot\vec{e}_{\varphi}(1)=\cos{(2\pi(i-1)/N)}$. Taking into account 
the relations
\begin{widetext}
\begin{eqnarray*}
\vec{r}_{i1}=R(\cos{(2\pi(i-1)/N)}-1,\sin{(2\pi(i-1)/N)}) \\
r_{i1}^{2}=4R^{2}\sin^{2}{(\pi(i-1)/N)} \\
\vec{r}_{i1}\cdot\vec{e}_{1}=R(t_{\varphi}\sin{(2\pi(i-1)/N)}+t_{r}(\cos{(2\pi(i-1)/N)}-1)) \\
\vec{r}_{i1}\cdot\vec{e}_{i}=R(t_{\varphi}\sin{(2\pi(i-1)/N)}+t_{r}(1-\cos{(2\pi(i-1)/N)})) \\
(\vec{r}_{i1}\cdot\vec{e}_{1})(\vec{r}_{i1}\cdot\vec{e}_{i})=R^{2}(\sin^{2}{(2\pi(i-1)/N)}-2t^{2}_{r}(1-\cos{(2\pi(i-1)/N)})) \, \, ,
\end{eqnarray*}
\end{widetext}
The dipolar energy per particle is given by
\begin{widetext}
\begin{equation}
U_{1}=-\frac{m^{2}}{8R^{3}}\sum^{N}_{i=2}\frac{1+\cos^{2}{(\pi(i-1)/N)}}{\sin^{3}{(\pi(i-1)/N)}}+\\
\frac{3t^{2}_{r}m^{2}}{16R^{3}}\sum^{N}_{i=2}\frac{1-\cos{(2\pi(i-1)/N)}}{\sin^{5}{(\pi(i-1)/N)}}   \, \, .
\label{Eq:9}
\end{equation}
\end{widetext}
Since $d=2R\sin{(\pi/N)}$,
the first part of $U_{1}$, $U_{10}$, coincides 
with a well-known expression used to describe
the dipolar energy of a ring~\cite{Mes14}:
\begin{equation}
U_{ring}=\frac{N}{2}U_{10}=\frac{m^{2}}{d^{3}}\frac{N}{2}\sin^{3}{(\pi/N)}\sum^{N}_{i=2}\frac{1+\cos^{2}{(\pi(i-1)/N)}}{\sin^{3}{(\pi(i-1)/N)}}
\label{Eq:10}
\end{equation}
The second part describes the increase of the dipolar 
energy due to the deviation of the direction
of the magnetic moment from the azimuthal direction,
which is supposed to be the same for all particles in the ring.
This energy per unit length is
\begin{widetext}
\begin{equation*}
e_{d}=-\frac{3M^{2}}{4d^{2}}\sin^{3}{(\pi/N)}\sum^{N}_{i=2}\frac{1-\cos{(2\pi(i-1)/N)}}{\sin^{5}{(\pi(i-1)/N)}}(\vec{e}\cdot\vec{t})^{2}=-\frac{3M^{2}K_{N}'}{d^{2}}(\vec{e}\cdot\vec{t})^{2}
\label{Eq:11}
\end{equation*}
\end{widetext}
where $\vec{e}$ are the unit vectors along  
the magnetic moments and $\vec{t}$
the tangent vectors to the ring all along its perimeter.
The value of $K_{N}'$ in this case is
\begin{equation}
K_{N}'=\frac{1}{4}\sin^{3}{(\pi/N)}\sum^{N}_{i=2}\frac{1-\cos{(2\pi(i-1)/N)}}{\sin^{5}{(\pi(i-1)/N)}}
\label{Eq:12}
\end{equation}
and it is numerically very close to $K_{N}$.

\section{Derivation of Eqs.1 and 2 in the text}

The energy of the dipolar interactions in a ring made
up of $N$ particles is
\begin{equation}
E_{d}=-\frac{\pi}{2}M_p^{2}\frac{\pi}{6}d^{3}NK_{N}'(\vec{e}\cdot\vec{t})^2 \, \, .
\label {Eq:1}
\end{equation}
Here, the angle between the vectors $\vec{e}$ and $\vec{t}$
is approximated to a constant value, 
assuming that it is the same all along the ring perimeter, and
$M_p$ is the magnetization of the particles with diameters $d$
($m=M_p \pi d^3/6$).
In order to calculate the energy of the ring,
we introduce the magnetization per unit length, $M=m/d$. Then the dipolar energy
per unit length $e_{d}=E_{d}/L$ ($N=L/d$) reads as
\begin{equation}
e_{d}=-3\frac{m^{2}}{d^{4}}K_{N}'(\vec{e}\cdot\vec{t})^{2} \, \, ,
\end{equation}
or in an equivalent form
\begin{equation}
e_{d}=-\frac{3K_{N}'}{d^{2}}M^{2}(\vec{e}\cdot\vec{t})^{2} \, \, .
\label{Eq:3}
\end{equation}
Consequently, the demagnetization factor $\Delta N$ of
a ring is $\Delta N=6K_{N}'/d^{2}$, and
the total magnetic energy per unit length of the ring then reads as
\begin{equation}
e_{m}=-M\vec{e}\cdot\vec{H}-\frac{1}{2}\Delta NM^{2}(\vec{e}\cdot\vec{t})^{2} \, \, .
\label{Eq:4}
\end{equation}
Assuming that the magnetic equilibrium is established faster than the mechanical equilibrium,
the governing equations for the magnetic equilibrium can be written as
\begin{equation}
\vec{K}_{\vec{e}}e_{m}=0 \, \, ,
\end{equation}
being $\vec{K}_{\vec{a}}=\vec{a}\times \frac{\partial}{\partial \vec{a}}$.
By considering an applied field with small amplitude, one can solve
this equation using the power series,
$\vec{e}=\vec{e}_{0}+\vec{e}_{1}+\vec{e}_{2}$.
The zero order solution is $\vec{e}_{0}=\vec{t}$,
and the condition  $\vec{e}^2=1$
gives $\vec{e}_0\cdot \vec{e}_{1}=0$,
and $\vec{e}_0\cdot \vec{e}_{2}=-\vec{e}_{1}^2/2$.
Up to the second order term $\vec{e}\cdot \vec{t}=1+\vec{e}_{2}\cdot \vec{e}_{t}$,
and the equation can be rewritten
after a straightforward calculation (see similar derivation in~\cite{Fer16}), as
\begin{equation}
\vec{e}_{1}=-\frac{MH}{\Delta NM^{2}}[\vec{t}\times[\vec{t}\times\vec{h}]] \, \, .
\label{Eq:5}
\end{equation}
The small amplitude hypothesis 
implies an applied field strength
lower than the dipolar field created by the magnetic particle, 
given by $H_{dip}=6 K_{N}' m/d^3$. Using the values of $m,d$
given in the manuscript, the dipolar field obtained is $H_{dip}\sim 0.6 {\rm Oe}$, 
which is actually larger than all the applied field strengths 
used in the experiments of ring deformation (Fig.2).

The torque per unit length can be written as,
$\vec{K}=-\vec{K}_{\vec{t}}e_{m}=\Delta N M^{2}(\vec{e}\cdot\vec{t})[\vec{t}\times\vec{e}]$.
Here the second order approximation gives
\begin{equation}
\vec{K}=\Delta NM^{2}[\vec{t}\times\vec{e}_{1}]+\Delta N M^{2}[\vec{t}\times\vec{e}_{2}] \, \, .
\label{Eq:6}
\end{equation}
The equation of magnetic equilibrium up to the second order terms
gives $-M[\vec{e}_{1}\times\vec{H}]-\Delta NM^{2}[\vec{e}_{2}\times\vec{t}]=0$,
and by using Eq.(\ref{Eq:5}) we obtain
\begin{equation}
\vec{K}=MH[\vec{t}\times\vec{h}]-\frac{H^{2}}{\Delta N}(\vec{t}\cdot\vec{h})[\vec{t}\times\vec{h}] \, \, .
\label{Eq:7}
\end{equation}
This corresponds to a linear energy density equal to
\begin{equation}
e_{r}=-MH\vec{t}\cdot\vec{h}+\frac{H^{2}}{2\Delta N}(\vec{t}\cdot\vec{h})^{2}  \, \, .
\label{Eq:7a}
\end{equation}

\section{Derivation of Eq. 8 in the text}

Let us start from the energy functional (Eq.7):
\begin{equation}
E=\frac{A}{2}\int{\Bigl( \frac{d\vartheta}{d l}\Bigl)^2 dl} +\frac{H^2}{2\Delta N}\int{\cos^2{\vartheta} dl}  \, \, ,
\label{Eq:10}
\end{equation}
where $\vec{t}=(\cos{\vartheta},\sin{\vartheta})$,
$d\vartheta/dl=-1/R$ is the curvature of the ring,
$l$ the natural parameter along the ring contour.
Minimizing the energy $\delta E = 0$
gives the Euler-Lagrange equation (Eq.8):
\begin{equation}
A\frac{d^2\vartheta}{dl^2}+\frac{H^2}{\Delta N}\cos{\vartheta}\sin{\vartheta}=0 \, \, ,
\label{Eq:11}
\end{equation}
where the angle $\theta$ changes from $0$
to $2\pi$
going around the ring with contour length $L$,
due to the close form of the ring, which is supposed to be inextensible.
Eq.~\ref{Eq:11} is equivalent to the dynamical system:
\begin{equation}
\frac{d \vartheta}{d l}=\omega \, ; \, \frac{d \omega}{d l}=-\frac{H^2}{\Delta N} \cos{(\vartheta)} \sin{(\vartheta)}\, \, ,
\label{Eq:12}
\end{equation}
which has the following fixed points,
$(\vartheta,\omega)=(0,0);(\pm \pi,0);(\pm \pi/2,0);(\pm 3\pi/2,0);...$
here, the first two are centers and second two are saddle points.
Eqs.\ref{Eq:12}
may be put in dimensionless form by introducing the
magnetoelastic number $C_m=\frac{H^2 L^2}{\Delta N A}$.
Fig.5 shows the phase portrait of the system described by Eqs.\ref{Eq:12}
for $C_m=5$.
There are two different phase trajectories. One corresponds to librations
around the centers with limited variation of the angle $\vartheta$,
\begin{figure}[t]
\begin{center}
\includegraphics[width=\columnwidth,keepaspectratio]{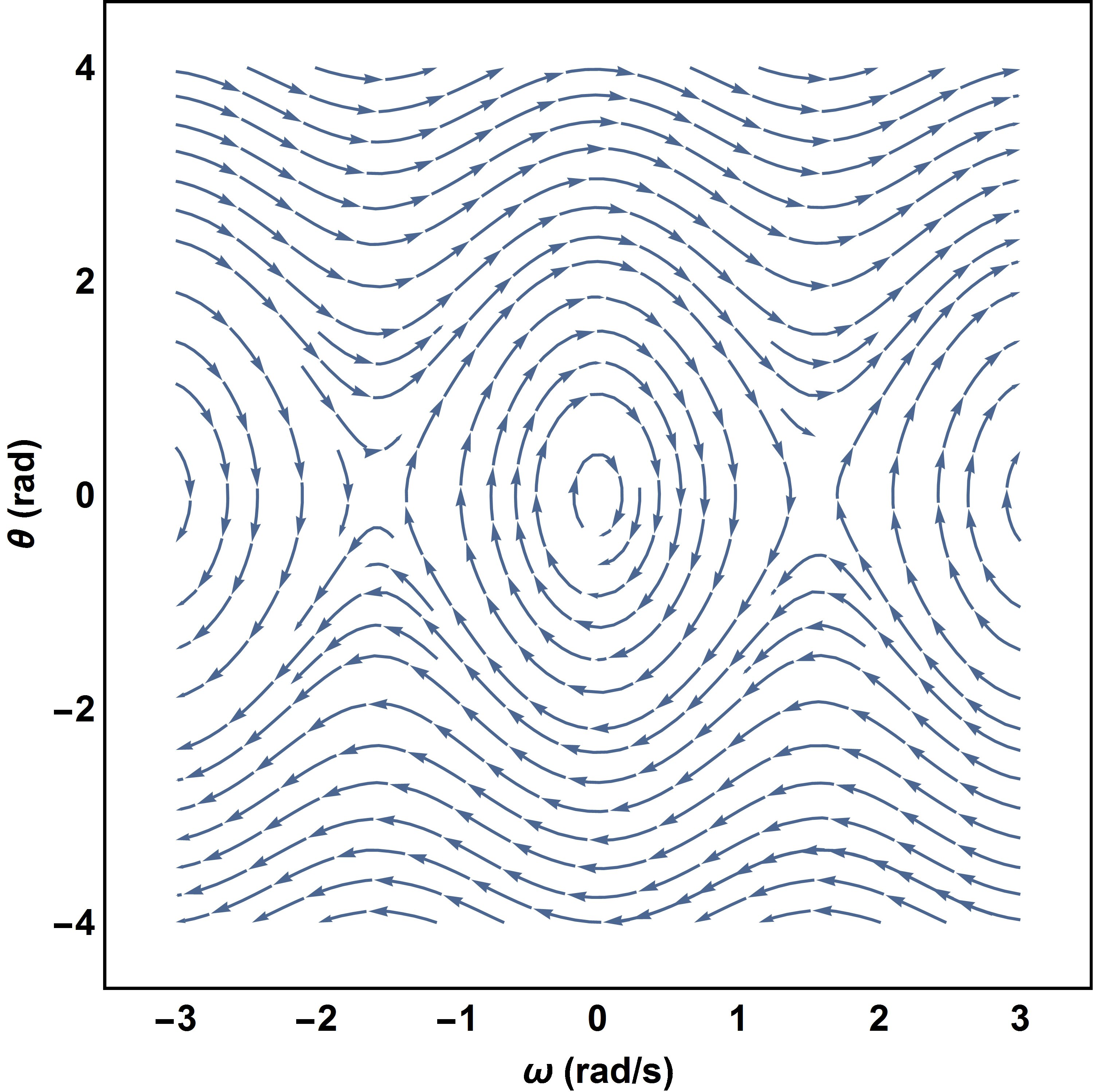}
\caption{Phase portrait of dynamical system given by Eq.~\ref{Eq:12} for $C_m=5$.}
\label{fig_5}
\end{center}
\end{figure}
while the second corresponds to the variation of the angle  $\vartheta$
by $2\pi$ along the contour. Only these trajectories correspond to
possible solutions. The first integral of
Eq.~\ref{Eq:11} is $C_1$ and is given by,
\begin{equation}
\frac{A}{2}\Big( \frac{d\vartheta }{dl} \Big)^2+ \frac{H^2}{2\Delta N}\sin^2{(\vartheta)}=C_1\, \, .
\label{Eq:13}
\end{equation}
Assuming a nonvanishing curvature of the ring, $C_1>\frac{H^2}{2\Delta N}$ and
\begin{equation}
\frac{d\vartheta}{dl} =\sqrt{\frac{2}{A}}  \sqrt{C_1-\frac{H^2}{2\Delta N}\sin^2{(\vartheta)}} \, \, .
\label{Eq:14}
\end{equation}
The solution of Eq.~\ref{Eq:14} reads as:
\begin{equation}
\int_0^{\vartheta}\frac{d\vartheta}{\sqrt{1-k^2\sin^2{(\vartheta)}}}=\sqrt{\frac{2C_1}{A}}l
\label{Eq:15}
\end{equation}
or
\begin{equation}
\vartheta=am \Big( \sqrt{\frac{2 C_1}{A}}l,k \Big)  \, \, ,
\label{Eq:16}
\end{equation}
where $am$ is the amplitude of elliptic integral and $k^2=\frac{H^2}{2\Delta N C_1}$.
Since a change of the angle $\vartheta$
from $0$ to $\pi/2$ corresponds to a contour length $L/4$,
the parameter $k$ at given magnetoelastic number
can be determined from the following equation:
\begin{equation}
\sqrt{k^2}K(k)= \frac{1}{4}\sqrt{C_m} \, \, .
\label{Eq:17}
\end{equation}
where $K(k)$ is a complete elliptic integral of first kind
and modulus $k$.
The shape of the ring can be found by solving the set of equations:
\begin{equation}
\frac{dx}{dl}=\cos{(\vartheta)} \, ; \, \frac{dy}{dl}=\cos{(\vartheta)}  \, .
\label{Eq:18}
\end{equation}
Scaling the contour length $l$ by the length of the ring $L$
and using some properties of the elliptic functions,
the previous equations may be written as:
\begin{equation}
\frac{dx}{dl}=cn\Bigl( \sqrt{\frac{C_m}{k^2}}l,k \Bigl); \, \,
\frac{dy}{dl}=sn\Bigl( \sqrt{\frac{C_m}{k^2}}l,k \Bigl)\, \, ,
\label{Eq:18}
\end{equation}
where
$cn(u;k)$ and $sn(u;k)$
are the Jacobi elliptic functions of
first and second kind.

\bibliography{biblio}
\end{document}